%Paper: hep-ph/9212295
%From: jenkins@sphal.UCSD.EDU (Elizabeth E. Jenkins)
%Date: Mon, 21 Dec 92 13:37:20 -0800

%
% If you have epsf.tex, uncomment the following line and the figures
% will be included automatically. Otherwise, you can print them out
% separately.
%
%\input epsf
%
%
\input harvmac
%
%%%%%%%%%%%%%%%%%%%%%%%%%%%%%%%%%%%%%%%%%%%%%%%%%%%%%%%%%%%%%%%%%%%%%%
%
%  UCSD macros to overwrite some of the definitions in harvmac.tex
%  (include after harvmac.tex)
%  last modified 4/92
%
%%%%%%%%%%%%%%%%%%%%%%%%%%%%%%%%%%%%%%%%%%%%%%%%%%%%%%%%%%%%%%%%%%%%%%%
%
% modify the output routine for the little format
%
\ifx\answ\bigans
\else
\output={
  \almostshipout{\leftline{\vbox{\pagebody\makefootline}}}\advancepageno
}
\fi
%
%
% address
%

%
% grant numbers
%

%
% preprint number
%
\def\UCSD#1#2{\noindent#1\hfill #2%
\bigskip\supereject\global\hsize=\hsbody%
\footline={\hss\tenrm\folio\hss}}% restores pagenumbers
%
% abstract
%
\def\abstract#1{\centerline{\bf Abstract}\nobreak\medskip\nobreak\par #1}
%
%
% titlefont
%
%
\edef\tfontsize{ scaled\magstep3}
 \tfontsize  \tfontsize
 \tfontsize \font\titlei=cmmi10 \tfontsize
\font\titleis=cmmi7 \tfontsize \font\titleiss=cmmi5 \tfontsize
\font\titlesy=cmsy10 \tfontsize \font\titlesys=cmsy7 \tfontsize
\font\titlesyss=cmsy5 \tfontsize  \tfontsize
\skewchar\titlei='177 \skewchar\titleis='177 \skewchar\titleiss='177
\skewchar\titlesy='60 \skewchar\titlesys='60 \skewchar\titlesyss='60
%
%\def\titlefont{\def\rm{\fam0\titlerm}% switch to title font
%\textfont0=\titlerm \scriptfont0=\titlerms \scriptscriptfont0=\titlermss
%\textfont1=\titlei \scriptfont1=\titleis \scriptscriptfont1=\titleiss
%\textfont2=\titlesy \scriptfont2=\titlesys \scriptscriptfont2=\titlesyss
%\textfont\itfam=\titleit \def\it{\fam\itfam\titleit}\rm}
%
%
% math symbols
%
%---------------------------------------------------------------------
%
\def\inv{^{\raise.15ex\hbox{${\scriptscriptstyle -}$}\kern-.05em 1}}
  %prime
\def\lbar{{\lower.35ex\hbox{$\mathchar'26$}\mkern-10mu\lambda}} %lambda bar

%
%
% various slashed symbols
%
%
\def\slash#1{\rlap{$#1$}/} % slashes a character
\def\dsl{\,\raise.15ex\hbox{/}\mkern-13.5mu D} %this one can be subscripted
\def\delsl{\raise.15ex\hbox{/}\kern-.57em\partial}
\def\Ksl{\hbox{/\kern-.6000em\rm K}}
\def\Asl{\hbox{/\kern-.6500em \rm A}}
\def\Dsl{\hbox{/\kern-.6000em\rm D}} %roman D
\def\Qsl{\hbox{/\kern-.6000em\rm Q}}
\def\gradsl{\hbox{/\kern-.6500em$\nabla$}}
%
% space and backspace in l mode
%
\def\lspace{\ifx\answ\bigans{}\else\qquad\fi}
\def\lbspace{\ifx\answ\bigans{}\else\hskip-.2in\fi} % $$\lbspace...$$
%
%     boxes an equation
%
\def\boxeqn#1{\vcenter{\vbox{\hrule\hbox{\vrule\kern3pt\vbox{\kern3pt
        \hbox{${\displaystyle #1}$}\kern3pt}\kern3pt\vrule}\hrule}}}
%
%     draw a little box (end of proof symbol)
%     e.g. \mbox{.1}{.1}
%
\def\mbox#1#2{\vcenter{\hrule \hbox{\vrule height#2in
\kern#1in \vrule} \hrule}}
%
%
%
%     curly letters
%
   %curly letters

   \def\CL{{\cal L}}
  \def\CO{{\cal O}} 
\def\CQ{{\cal Q}}

%
%
%
%     derivatives
%
%

%

\def\bar#1{\overline{#1}}

\def\darr#1{\raise1.5ex\hbox{$\leftrightarrow$}\mkern-16.5mu #1}

%
 %pound sterling
%
\def\half{{\textstyle{1\over2}}} %puts a small half in a displayed eqn
\def\frac#1#2{{\textstyle{#1\over #2}}} %puts a small fraction
%in a displayed eqn
%
%
%     various math operators
%
%

\def\Tr{\mathop{\rm Tr}}

%
%
%
%

%
%       relations
%
\def\ltap{\ \raise.3ex\hbox{$<$\kern-.75em\lower1ex\hbox{$\sim$}}\ }
\def\gtap{\ \raise.3ex\hbox{$>$\kern-.75em\lower1ex\hbox{$\sim$}}\ }
\def\gl{\ \raise.5ex\hbox{$>$}\kern-.8em\lower.5ex\hbox{$<$}\ }
\def\roughly#1{\raise.3ex\hbox{$#1$\kern-.75em\lower1ex\hbox{$\sim$}}}
%
%
%       This defines et al., i.e., e.g., cf., etc.

%

%
\def\np#1#2#3{Nucl. Phys. B{#1} (#2) #3}
\def\pl#1#2#3{Phys. Lett. {#1}B (#2) #3}
\def\prl#1#2#3{Phys. Rev. Lett. {#1} (#2) #3}
\def\physrev#1#2#3{Phys. Rev. {#1} (#2) #3}

\relax
\ifx\epsfbox\notincluded\message{(NO epsf.tex, FIGURES WILL BE IGNORED)}
\def\insertfig#1{}% null macro
\else\message{(FIGURES WILL BE INCLUDED)}
\def\insertfig#1{
\midinsert\centerline{\epsfxsize=4truein
\epsfbox{#1}}\endinsert}
\fi
\noblackbox
\def\lcsb{\Lambda_{\chi}}
\def\aem{\alpha_{em}\,}
\def\m{M_\xi}
\def\q{\CQ_\xi}
\def\asl{\hbox{/\kern-.6500em A}}
\def\s{{\slash s}}
\def\hyper#1{{\Delta^{(#1)}}}
\def\hyperem#1{{\Delta_{em}^{(#1)}}}
\vskip 1.in
\centerline{{\titlefont{Heavy Meson Masses}}}
\medskip
\centerline{{\titlefont{in Chiral Perturbation Theory}}}
\medskip
\centerline{{\titlefont{With Heavy Quark Symmetry}}}
\vskip .2in
\centerline{Elizabeth Jenkins\footnote{${}^*$}{On leave from the
University of California at San Diego.} }
\medskip
\centerline{\sl CERN TH Division, CH-1211 Geneva 23, Switzerland}
\vfill
\abstract{
The $SU(3)$ and hyperfine mass splittings of mesons
containing a single heavy quark are computed to
one-loop order in chiral perturbation theory
with heavy quark spin symmetry.  Electromagnetic
contributions of order $\aem$ are included.
The observed values of the mass splittings are
consistent with the one-loop chiral perturbation
theory calculation.  The $d-u$ hyperfine
splitting
$(D^{* +} - D^+) - (D^{*0} - D^0)$ is equal
to $(m_d - m_u)/ m_s$ times the $s-d$ hyperfine
splitting
$(D_s^{* +} - D_s^+) - (D^{*+} - D^+)$ upto
electromagnetic contributions.  The small
observed value of the $s-d$ hyperfine
splitting implies that the $d-u$ hyperfine
splitting is completely dominated by
electromagnetic mass contributions.
}
\vfill
\UCSD{\vbox{
\hbox{CERN-TH.6765/92}\vskip-0.1truecm
\hbox{hep-ph/9212295}}}{December 1992}
%\draftmode
\eject

\newsec{Introduction}

The spectrum of masses for the lowest-lying
pseudoscalar and vector mesons which contain
a single heavy quark has been of recent
theoretical interest.  Two approximate
symmetries of QCD, chiral symmetry for the
light $u, d$ and $s$ quarks and heavy quark
spin-flavor symmetry
\ref\iswis{N. Isgur and M.B. Wise, \pl {232}
{1989}{113}\semi \pl {237}{1990}{527}  }
for the heavy $c$ and
$b$ quarks, can be exploited to calculate
the heavy meson masses.  These symmetries
have led to a number of theoretical
predictions for the heavy meson masses.
The hyperfine mass splittings $D^* -D$
and $B^* -B$ are related using
heavy quark spin-flavor symmetry.  Corrections
to the naive $1/m_Q$ scaling of these
splittings can be calculated in HQET
\ref\falk{A.F. Falk, B. Grinstein and M.E. Luke,
\np{357}{1991}{185} }.
The $SU(3)$ flavor splitting $D_s -D$ can be
calculated using chiral Lagrangian techniques.
A leading nonanalytic chiral correction
to this mass splitting is calculated in
ref.~\ref\goity{J.L. Goity, CEBAF-TH-92-16
[{\tt hep-ph/9206230}] }.
Rosner and Wise
\ref\roswis{J.L. Rosner and M.B. Wise, CALT-68-1807 }
recently analyzed
the heavy meson mass spectrum to first order
in light quark masses and to order $1/ m_Q$.
The leading effects from
electromagnetism proportional to one power of
$\aem$ or $\aem / m_Q$ also were considered.
Rosner and Wise obtain the mass relation
$\left\{ (B_s^* - B_s) - (B^{*0} - B^0)
\right\} = (m_c / m_b) \left\{ (D_s^* - D_s)
- (D^{*+} - D^+) \right\}$.
The electromagnetic splittings
of the mesons have been estimated
using dispersion relation techniques
\ref\goitytwo{J.L. Goity, CEBAF-TH-92-26
[{\tt hep-ph/9212230}] }.
Estimates of the hyperfine mass
splittings for the $D$ and $B$ mesons are
discussed in
ref.~\ref\randall{L. Randall and E. Sather,
MIT-CTP\#2166 }.

In this work, the masses of the
$D$, $D^*$, $B$ and $B^*$ mesons are calculated
to one-loop order in chiral perturbation theory
using a chiral Lagrangian which incorporates
heavy quark spin-flavor symmetry and its
breaking.  The calculation naturally breaks up
into the calculation of
$SU(3)$ mass splittings
which preserve heavy quark spin symmetry
and
the calculation of
hyperfine mass splittings which explicitly
violate heavy quark spin symmetry.  The
one-loop chiral perturbation theory calculation
results in contributions to the meson masses
which are nonanalytic in the light quark masses.
The leading contribution to the $s-d$
hyperfine mass difference
$(D_s^{* +} - D_s^+) - (D^{*+} - D^+)$
is nonanalytic in the light quark masses.  The
small measured value of this hyperfine mass
difference implies nearly exact cancellation
between the nonanalytic contribution and
a one-loop counterterm.  The contribution to
the $d-u$ hyperfine mass splitting
$(D^{* +} - D^+) - (D^{*0} - D^0)$
due to light quark mass differences is
simply equal to the $s-d$ mass splitting
times the ratio
$(m_s - m_d)/(m_d - m_u)$.
Experimentally, the hyperfine mass
differences do not satisfy such a relation;
both hyperfine mass splittings are of order one
MeV.  Thus, the $d-u$ hyperfine mass difference
must be dominated by electromagnetic effects.
(There is no electromagnetic contribution to
the $s-d$ hyperfine mass difference since the
$s$ and $d$ quarks have equal charges.)
Finally, nonanalytic corrections do not
violate the Rosner-Wise mass relation.  The
relation is violated at order $(1/ m_Q)^2$
in the heavy quark mass expansion.

The organization of this paper is as follows.
The chiral Lagrangian for
heavy mesons is constructed in Sect.~2.  The leading order Lagrangian
for heavy meson-pion interactions which
satisfies heavy quark spin-flavor symmetry
is reviewed.  Higher order terms in the
chiral Lagrangian which are relevant for the
calculation of the heavy meson mass
splittings
are discussed next.  These terms break
chiral symmetry and heavy quark symmetry.
The terms are classified by
the number of insertions
of the light quark mass matrix they contain
and by whether or not they violate the heavy quark
spin symmetry.
Finally, chiral Lagrangian terms describing
the effects of electromagnetism on the
heavy meson masses are detailed.  Only the
leading operators which are suppressed by
a single power of $\aem$ are included.
Electromagnetic effects break chiral symmetry
and heavy quark spin-flavor symmetry since
the charges and magnetic moments of the light and
heavy quarks violate these symmetries.
Sect.~3 contains the calculation of the
$SU(3)$ and hyperfine mass splittings
of the heavy mesons to one-loop order
in chiral perturbation theory
neglecting the masses
of the $u$ and $d$ quarks and electromagnetic
effects.  Sect.~4 generalizes the calculation
to include nonvanishing $u$ and $d$ quark
masses, as well as electromagnetic corrections.
In Sect.~5, the
theoretical formul\ae\ for the mass
splittings are compared with experiment.
Sect.~6 contains final remarks.

\newsec{Chiral Lagrangian for Heavy Mesons}

The lowest lying mesons which contain a single heavy quark
$Q$ are pseudoscalar and vector mesons with spin of
the light degrees of freedom $s_\ell = \frac 1 2$.
Degenerate spin zero and spin one mesons result when
the spin of the light degrees of freedom is combined with
the spin of the heavy quark $s_Q = \frac 1 2$.
Since the pseudoscalar and vector mesons are
degenerate in the heavy quark mass limit and the
spin of the heavy quark is conserved by low energy
strong interactions, it is convenient
to introduce a single field
\ref\hgtasi{H. Georgi, {\it Heavy Quark Effective Field
Theory,} in Proceedings of the Theoretical Advanced Study
Institute 1991, ed. R.K. Ellis, C.T. Hill and J.D. Lykken,
World Scientific (1992) }
for the heavy meson multiplet
\foot{The usual definition of the $4 \times 4$ Dirac matrix
$H^{(Q)}_a$ does not include the projector
${ {(1- \slash v)} / 2 }$.
This projector, however, results from commuting the heavy
quark velocity projector
${ {(1+ \slash v)} / 2 }$ through the matrix in
square brackets since
$v \cdot P^* =0$. }
,
\eqn\hmatrix{
H^{(Q)}_a = { {(1+ \slash v)} \over 2 }
\left[ P_{a \mu}^{*(Q)} \gamma^\mu - P_a^{(Q)}
\gamma_5 \right] { {(1- \slash v)} \over 2 }  ,
}
where
$P_a^{(Q)}$ and $P_a^{*(Q)}$ are the pseudoscalar
and vector mesons fields, respectively, with
flavor quantum numbers $Q \bar q_a$ where $Q = c, b$ and
$a=1, 2, 3$ (or $u, d, s$).
For $Q = c$, these fields are the
$(D_1, D_2, D_3) = (D^0, D^+, D_s)$ and
$(D^*_1, D^*_2, D^*_3) = (D^{*0}, D^{*+}, D^*_s )$
mesons, whereas for $Q = b$, they are the
$(B_1, B_2, B_3) = (B^-, \bar B^0, \bar B_s)$ and
$(B^*_1, B^*_2, B^*_3) = (B^{*-}, \bar B^{*0}, \bar B^*_s )$
mesons.
The heavy meson field \hmatrix\
also carries a velocity subscript
which has been suppressed.
The velocity-dependent field
\ref\georgi{H. Georgi, \pl{240}{1990}{447} }
is defined by the field redefinition
\eqn\velfield{
H_v^{(Q)} = \sqrt{m_H} \ \, e^{i \, m_H \slash v \, v_{\mu}
x^{\mu} }  H^{(Q)}(x)  ,
}
which removes the large mass
$m_H \Tr \bar H H$ from the HQET Lagrangian.
The velocity-dependent field satisfies $\slash v H_v = H_v$
and $H_v \slash v = - H_v$.
The conjugate heavy meson field is defined by
\eqn\hconj{
\bar H^{a (Q)} = \gamma^0 H_a^{(Q) \dagger} \gamma^0
= { {(1- \slash v)} \over 2 }
\left[ P^{*a (Q) \dagger}_\mu \gamma^{\mu} +
P^{a (Q) \dagger} \gamma_5 \right]
{ {(1+ \slash v)} \over 2 }  .
}

The low energy interactions of mesons containing
a single heavy quark with pions are described
by a chiral Lagrangian which respects heavy quark
spin-flavor symmetry
\ref\wise{M.B. Wise, \physrev{D45}{1992}{2188} }
\ref\bd{G. Burdman and J.F. Donoghue, \pl{280}{1992}{287} }
\ref\tai{T.M. Yan, H.Y. Cheng, C.Y. Cheung, G.L. Lin,
Y.C. Lin and H.L. Yu, \physrev{D46}{1992}{1148} }.
The heavy meson field $H_v$ transforms as a doublet under
heavy quark spin symmetry and as a $\bar 3$ under
flavor $SU(3)_V$,
\eqn\htran{
H^{(Q)}_{a} \rightarrow S_{Q} \left(
H^{(Q)} U^\dagger \right)_a  \qquad .
}
In addition, the heavy meson field transforms as a doublet
under heavy quark flavor symmetry.
The field redefinition which defines the
velocity-dependent field $H_v$
removes the singlet mass term, the mass which
is invariant under chiral symmetry and
heavy quark spin-flavor symmetry,
from the chiral Lagrangian.  The chiral Lagrangian
which is formulated using this velocity-dependent field
has a valid derivative expansion and loop expansion
because the chiral Lagrangian
has a systematic expansion in $1/m_H$ (or $1/ m_Q$)
\ref\jmone{E. Jenkins and A.V. Manohar, \pl{255}{1991}{558} }
\ref\jmtwo{E. Jenkins and A.V. Manohar, {\sl Baryon Chiral
Perturbation Theory}, in Proceedings of the Workshop on
``Effective Field Theories of the Standard Model,''
ed. U. Meissner, World Scientific (1992) }.
Derivatives acting on the heavy meson field produce
factors of $k$, the residual off-shell momentum of the
heavy meson field, or equivalently, the typical
momentum of the pions interacting with the
heavy mesons.  Higher dimensional terms of the chiral
Lagrangian are suppressed by powers of either $1/ m_H$ or
$1/ \lcsb$, where $\lcsb \sim 1$~GeV is the chiral symmetry
breaking scale \ref\am{A.V. Manohar and H. Georgi,
\np{234}{1984}{189}}.  Note that the value of $m_H$ used
to define velocity-dependent heavy meson fields
will be quite different for the $D^{(*)}$ and
$B^{(*)}$ mesons.  The mass $m_H$ can be expanded
in a power series in $m_Q$.  To leading order in
this expansion, $m_H = m_Q$.

The leading order chiral Lagrangian for heavy mesons
interacting with pions is given by
\eqn\lvzero{
\CL_v^0 = -i \Tr \bar H_v (v \cdot \partial) H_v
+i \Tr \bar H_v H_v (v \cdot V)
+2g \Tr \bar H_v H_v ( S_{\ell \, v} \cdot A ) ,
}
where the heavy and light quark flavor quantum numbers
of the heavy meson field have been suppressed.
In eq.~\lvzero, all
traces are taken over Dirac spinor indices,
light quark $SU(3)_V$ flavor indices $a= u,d,s,$ and
heavy quark flavor indices $Q= c, b$.  The
pion octet
\eqn\pion{
\pi = {1 \over {\sqrt{2}}} \pmatrix{ {1\over\sqrt2}\pi^0 +
{1\over\sqrt6}\eta&
\pi^+ & K^+\cr
\pi^-& -{1\over\sqrt2}\pi^0 + {1\over\sqrt6}\eta&K^0\cr
K^- &\bar K^0 &- {2\over\sqrt6}\eta\cr},
}
appears in Lagrangian~$\CL_v^0$ through the
vector and axial vector linear combinations
\eqn\av{
V^{\mu} = \frac 1 2 ( \xi \partial^{\mu} \xi^{\dagger}
+ \xi^\dagger \partial^{\mu} \xi ) , \qquad
A^{\mu} = \frac i 2 (\xi \partial^{\mu} \xi^{\dagger}
- \xi^\dagger \partial^{\mu} \xi ),
}
where
\eqn\sigmaxi{
\xi = e^{i\pi/f}, \quad \Sigma = \xi^2 = e^{2 i \pi/f},
}
and $f \approx 93$~MeV is the pion decay constant.
Under $SU(3)_L \times SU(3)_R$ chiral symmetry,
\eqn\tran{\eqalign{
&\Sigma \rightarrow L \Sigma R^\dagger, \qquad
\xi \rightarrow L \xi U^\dagger = U \xi R^\dagger, \cr
}}
where $U$ is defined by the transformation of $\xi$.
The Dirac structure of the heavy meson chiral
Lagrangian has been replaced by the velocity vector
$v^{\mu}$ and the velocity-dependent spin operator
$S_{\ell\, v}^{\mu}$ for the light degrees of freedom
\jmone.
The spin operator for the light degrees of freedom
is defined by
$\Tr \bar H_v H_v \gamma^{\mu} \gamma_5 \equiv
2 \Tr \bar H_v H_v S_{\ell\, v}^{\mu} \ ,$
and satisfies the identity
$\left( v \cdot S_{\ell \, v} \right)=0 $.
In the
rest frame $v^{\mu} = (1 , \vec 0)$ of the heavy meson,
$S_{\ell}^{\mu}  = ( 0, {\vec \sigma}/ 2 )$.
The axial vector combination of pions
$A^{\mu}= \partial^{\mu} \pi / f + ...$ couples to the
spin of the light degrees of freedom in the heavy
mesons.
The spin of the
heavy quark is unaffected by the low energy pion
interactions of the heavy mesons, however, since
an axial vector coupling to the spin of the
heavy quark $\Tr \bar H_v \asl \gamma_5 H_v  \equiv
2 \Tr \bar H_v (S_{Q\, v} \cdot A ) H_v $
is forbidden by heavy quark spin symmetry.
Thus, the chiral Lagrangian
$\CL_v^0$ depends on only a single axial vector
coupling constant $g$.  The coupling $g$ is constrained
by the radiative and pion decay widths of the vector
mesons, $g^2 \ltap 0.5$
\ref\accmor{The ACCMOR Collaboration (S. Barlag {\it et al.}),
\pl {278}{1992}{480} }
\ref\cleoone{The CLEO Collaboration (F. Butler {\it et al.}),
Phys. Rev. Lett. 69 (1992) 2041 }
\ref\drad{J.F. Amundson, C.G. Boyd, E. Jenkins, M. Luke,
A.V. Manohar, J.L. Rosner, M.J. Savage and M.B. Wise,
[{\tt hep-ph/9209241}] \semi
P. Cho and H. Georgi, HUTP-92/A043 [{\tt hep-ph/9209239}]\semi
H.Y. Cheng, C.Y. Cheung, G.L. Lin, Y.C. Lin, T.M. Yan
and H.L. Yu, CLNS 92/1158 }.
The propagator of the $H_v$ field is easily determined
from the leading order chiral Lagrangian to be
$-i/(k \cdot v)$.  Note that
$H_v$ is normalized to unity which implies that
the fields $P$ and $P^*$ are normalized to $2$,
since
$\Tr \bar H_v H_v = 2 ( P_{a \mu}^{*\dagger} P_a^{* \mu}
-P_a^\dagger P_a )$.  Loop calculations of corrections
which respect heavy quark spin symmetry can be calculated
using the field $H_v$ \ref\chotwo{P. Cho, HUTP-92/A039 [{\tt
hep-ph/9208244}] }.
Heavy quark spin symmetry violating corrections are best
calculated using the fields $P$ and $P^*$.

Higher dimensional operators of the chiral Lagrangian which
break heavy quark spin-flavor symmetry and chiral
symmetry involve factors of $1/ m_Q$ or insertions of the
light quark mass matrix $M = {\rm diag}(m_u, m_d, m_s)$.
For the calculation of the heavy meson masses, it is
useful to classify the symmetry-violating operators by
the number of insertions of
the quark mass matrix and whether or not they violate the
heavy quark spin symmetry. Operators which respect heavy
quark spin symmetry have coefficients which start at
$\CO(1)$ in the $1/m_Q$ expansion, whereas operators which
violate heavy quark spin symmetry have coefficients which
start at $\CO(1/m_Q)$. Counterterms to the one-loop calculation of the
heavy meson masses  are proportional to two powers of the light quark
masses. Thus terms with up to
two insertions of $M$ are needed for the one-loop calculation.

The singlet mass term $m_H$ is a function of $1/m_Q$, but is
irrelevant for the calculation of meson mass splittings.
There is only one other term with no insertions of the
light quark mass matrix,
\eqn\lvonem{ \CL_v^{\s}  = -{\Delta\over8}
\Tr \bar H_v \sigma^{\mu
\nu} H_v \sigma_{\mu \nu} =-\Delta
\Tr \bar H_v S_{Q\, v}^{\alpha} H_v S_{\ell \, v \, \alpha}.
}
This term violates heavy quark spin-flavor symmetry and
is responsible for the hyperfine $(P^* - P)$ mass splitting
at leading order. The parameter $\Delta$ is a function of $1/m_Q$
which starts at linear order in $1/m_Q$, because violations of
heavy quark spin symmetry are suppressed by at least one power of
$1/m_Q$.
The value of
$
-S_Q\cdot S_\ell=\vec S_Q \cdot \vec S_\ell = \frac 1 2 \left( \vec S_T^2
- \vec S_Q^2 -\vec S_\ell^2 \right)
$
equals
$- \frac 3 4$ for pseudoscalar mesons
$P$ and $\frac 1 4$ for vector mesons $P^*$.  Thus,
this operator results in a hyperfine
splitting $(P^* - P) = \Delta$ at tree level.
Note that the Dirac tensor structure
$\Tr \bar H_v \sigma^{\mu \nu} H_v \sigma_{\mu \nu} $
is the only possible spin symmetry-violating Lorentz
invariant since the operators
$\Tr \bar H_v \sigma^{\mu \nu} \sigma_{\mu \nu} H_v $
and
$\Tr \bar H_v H_v \sigma^{\mu \nu} \sigma_{\mu \nu} $
are proportional to $\Tr \bar H_v H_v$ because
$\vec S_Q^2$ and $\vec S_\ell^2$ are
constants for the heavy meson field.

The terms in the chiral Lagrangian which
are proportional to a light quark mass and which
respect the
heavy quark spin symmetry are given by
\eqn\lvm{
\CL_v^M =
a \Tr \bar H_v H_v \m
+ \sigma \Tr \m \Tr \bar H_v H_v \ ,
}
where

$a$ and $\sigma$ are functions of $1/m_Q$ which start
at $\CO(1)$.
The term
proportional to $a$ results in $SU(3)_V$-violating
mass splittings amongst the $P^{(*)}_a$ mesons.
The term proportional to $\sigma$ leads to a singlet
contribution to the masses which depends linearly
on the light quark masses, and is the heavy meson analog
of the pion-nucleon sigma term.

Chiral Lagrangian terms with one insertion of the light
quark mass matrix which violate the heavy quark spin symmetry
are
\eqn\lvmonem{\CL_v^{M \s}  =
-\frac18\hyper a \Tr \bar H_v \sigma^{\mu
\nu} H_v \sigma_{\mu \nu} \m
-\frac18\hyper\sigma \Tr \m
\ \Tr \bar H_v
\sigma^{\mu \nu} H_v \sigma_{\mu \nu} \ .
}
These two terms
produce hyperfine splittings proportional to
a light quark mass.  The term proportional to
$\hyper a$ leads to light quark flavor-dependent
hyperfine splittings,
whereas the term proportional
to $\hyper\sigma$ yields a flavor singlet contribution to the
hyperfine splittings
which is linear in the light quark masses.

Chiral Lagrangian terms with two insertions of the
light quark mass matrix which preserve and violate the
heavy quark spin symmetry are
\eqn\lvmm{
\CL_v^{M M} =
b \Tr \bar H_v H_v \m \m
+ c \Tr \m
\ \Tr \bar H_v H_v \m
+ d \Tr \m \m
\ \Tr \bar H_v H_v \ ,
}
and
\eqn\lvmmonem{\eqalign{
\CL_v^{M M \s} =&
-\frac18\hyper b
\Tr \bar H_v \sigma^{\mu \nu} H_v \sigma_{\mu \nu}
\m \m
\cr
&-\frac18 \hyper c \Tr \m
\ \Tr \bar H_v \sigma^{\mu \nu} H_v \sigma_{\mu \nu}
\m
\cr
&-\frac18 \hyper d\Tr \m \m
\ \Tr \bar H_v \sigma^{\mu \nu} H_v \sigma_{\mu \nu}
\ ,
\cr
}}
respectively. The coefficients $b$, $c$ and $d$ are functions of
$1/m_Q$ which begin at $\CO(1)$, whereas the coefficients
$\hyper b$, $\hyper c$, and $\hyper d$ are functions of
$1/m_Q$ which begin at $\CO(1/m_Q)$.\foot{
The coefficients which
violate heavy quark spin symmetry
({\it i.e.} $\Delta$,
$\hyper \sigma$, $\hyper a$, $\hyper b$,
$\hyper c$, and $\hyper d$) also have calculable
logarithmic dependence on the heavy quark mass from
perturbative QCD \falk.}

Electromagnetic mass splittings are incorporated
into the chiral
Lagrangian framework by including terms which
depend quadratically on the charges of the heavy
and light quarks, and are suppressed by at least
one power of $\aem$.
The heavy quark charges
$q_c = \frac 2 3$, $q_b = -\frac 1 3$
respect chiral $SU(3)_V$ symmetry but break
heavy quark flavor symmetry.  The light quark
charges appear in the chiral Lagrangian through
the light quark charge matrix
$\CQ = {\rm diag}\left( \frac 2 3 , - \frac 1 3,
-\frac 1 3 \right)$ which transforms nontrivially
under $SU(3)_V$.  The light quark charge matrix
appears in operators of the chiral
Lagrangian in the linear combination
$\q=\frac 1 2
\left(\xi \CQ \xi^\dagger
+ \xi^\dagger \CQ \xi \right)$, which transforms
simply under chiral symmetry, $\q\rightarrow U\q U^\dagger$.
In the following, terms of order $\aem$ with insertions of
the quark
mass matrix are neglected. The number of chiral invariants
constructed with $\q$ is less than the number for $\m$, since
$\Tr \q = \frac 1 2
\Tr \left( \xi \CQ \xi^\dagger
+ \xi^\dagger \CQ \xi \right)
= \Tr \CQ = 0$.

Heavy quark flavor symmetry-violating but heavy quark
spin symmetry-conserving
contributions to the $SU(3)$-invariant mass of
the heavy meson multiplet are generated by
the operator $\aem\,q_Q^2 \Tr \bar H_v H_v$
which contains two factors of the heavy quark charge. This
term does not contribute to the meson mass splittings,
however, and need not be considered here.
In addition, there is a term with two factors of the heavy
quark charge which violates the heavy quark spin and flavor symmetries,
\eqn\lqqonem{ \CL_v^{q_Q q_Q \s} =
-{1\over8}\Delta_{em}\aem \Tr \bar H_v q_Q^2 \sigma^{\mu
\nu} H_v \sigma_{\mu \nu} , }
and which contributes differently to the $SU(3)_V$ invariant
hyperfine splittings for the $D$ and $B$ mesons.
Electromagnetic contributions from terms
containing one factor of the light
quark charge matrix and one factor of the heavy quark
charge are given by
\eqn\lqbigQ{ \CL_v^{q_Q Q} =  a_{em}\aem \Tr \bar
H_v q_Q H_v \q  ,}
and
\eqn\lvqbigQonem{
\CL_v^{q_Q \CQ \s} =
-\frac18\hyperem a\ \aem
\Tr \bar H_v \,q_Q \sigma^{\mu \nu} H_v
\q
\sigma_{\mu \nu} ,
}
which preserve and violate the heavy quark spin symmetry,
respectively.
The terms in $\CL_v^{q_Q Q}$ and $\CL_v^{q_Q Q \s}$
violate heavy quark
flavor symmetry and
isospin.
Terms in  the chiral
Lagrangian with two insertions of the light quark
mass matrix are given by
\eqn\lvbigQbigQ{
\CL_v^{\CQ \CQ} =
b_{em} \aem\Tr \bar H_v H_v \q \q
+  d_{em} \aem\Tr \q \q
\ \Tr \bar H_v H_v,
}
which respects heavy quark symmetry, and by
\eqn\lvbigqbigqonem{\eqalign{
\CL_v^{Q Q \s} =&
-\frac18\hyperem b\ \aem
\Tr \bar H_v \sigma^{\mu \nu} H_v \sigma_{\mu \nu}
\q \q
\cr
&-\frac18\hyperem d\ \aem
\Tr \q \q
\ \Tr \bar H_v \sigma^{\mu \nu} H_v \sigma_{\mu \nu},
\cr
}}
which violates heavy quark spin symmetry.
The first terms in $\CL_v^{Q Q}$ and
$\CL_v^{Q Q \s}$ violate isospin.

The electromagnetic mass terms eq.~\lqbigQ\ and eq.~\lvbigQbigQ\
are expected to be of order 2--3~MeV, which is the size of typical
electromagnetic contributions to hadron masses. The electromagnetic
hyperfine mass terms eqs.~\lqqonem, \lvqbigQonem\ and~\lvbigqbigqonem\ are
due to the electromagnetic
hyperfine interaction between the light and heavy quarks, and are
expected to be of order $\Lambda/m_Q$ times the spin-independent
electromagnetic splitting.  Here $\Lambda$ denotes a
typical hadronic scale, which in a quark model is of order a
constituent
quark mass of 350--550~MeV.  Thus, eq.~\lqqonem,\lvqbigQonem\
and \lvbigqbigqonem\ each contribute
0.5--1~MeV to the $D$ meson masses and $0.2-0.3$~MeV to the
$B$ meson masses. Electromagnetic contributions
are relevant only for the isospin and isospin hyperfine
splittings of the heavy mesons.

\newsec{Calculation of Heavy Meson Masses
Neglecting $m_u$ and $m_d$}

In this section, the heavy meson masses are calculated
to one-loop order in chiral perturbation theory
with $m_u = m_d = 0$ and $\aem = 0$.  In this limit,
isospin is an exact symmetry, and electromagnetic
contributions to the masses are ignored.  Since the
strange quark is much heavier than the $u$ and $d$
quarks and $m_s - m_u \approx m_s - m_d \approx m_s$,
$SU(3)_V$ splittings can be determined by neglecting
the $u$ and $d$ quark masses relative to the strange
quark mass.  Isospin splittings will be considered
in the next section where the calculation is
generalized to include nonvanishing $u$ and $d$
quark masses.
\foot{Electromagnetic effects do not contribute to $SU(3)_V$
breaking mass splittings amongst mesons carrying
$d$ and $s$ light flavor quantum numbers since the
charges of the $d$ and $s$ quarks are the same.  Thus,
if one restricts the formul\ae\ given in this section to
these mesons, all $SU(3)_V$-violating contributions
are taken into account.  Neglected electromagnetic
contributions still result in small heavy quark flavor symmetry
violation, however.}

The one-loop calculation of the heavy meson masses naturally
separates into one-loop calculations for two different
linear combinations of the masses.  The first linear
combination
$\frac 1 4 \left( P_a+ 3P^*_a \right)$ respects heavy
quark spin symmetry and yields the purely
$SU(3)_V$-symmetric
and $SU(3)_V$-violating contributions to the
meson masses.  The second linear combination
$\left( P^*_a- P_a \right)$ violates heavy quark spin
symmetry and yields the $SU(3)_V$-symmetric and
$SU(3)_V$-violating contributions to the hyperfine mass
splittings of the heavy mesons.

To one-loop,
the chiral expansions for these two linear combinations
of the heavy meson masses can be written in
the form\foot{Analogous formul\ae\ are obtained for the
baryon octet and decuplet masses
\ref\gassleut{J. Gasser and H. Leutwyler,
Phys. Rep. 87 (1982) 77}
\ref\j{E. Jenkins, Nucl. Phys. B368 (1992) 190}
\jmtwo.  }
\eqn\massave{
\frac 1 4 \left( P_a+ 3P^*_a \right) = m_H + \alpha_a
- \beta_a { M_K^3 \over {16 \pi f^2 }}
+ \left( \gamma_a - \lambda_a\, \alpha_a \right)
{M_K^2 \over {16 \pi^2 f^2}} \ln \left( M_K^2 / \mu^2
\right) + c_a ,
}
and
\eqn\massdiff{
\left( P^*_a- P_a \right) = \Delta
+ \left( \gamma_a - \lambda_a\, \Delta \right)
{M_K^2 \over {16 \pi^2 f^2}} \ln \left( M_K^2 / \mu^2
\right) + \delta c_a ,
}
where $\alpha_a$, the tree-level mass splittings
calculated from $\CL_v^M$, are shown in
\fig\figone{Tree-level heavy meson mass splittings arising
from $\CL_v^M$ and $\CL_v^{\s}$.  Solid lines
denote the $H_v$ field.  The solid triangle (a)
denotes the $SU(3)_V$-violating vertex in $\CL_v^M$
proportional to $a$.  The vertex results in a
$\left( P_s^{(*)} - P^{(*)} \right)$ mass splitting
which is spin independent.  The solid diamond (b)
represents the vertex proportional to $\sigma$ in $\CL_v^M$
which yields a singlet contribution to the pseudoscalar
and vector meson masses.  The solid square vertex (c)
represents the vertex in $\CL_v^{\s}$.
The vertex leads to a $SU(3)_V$ flavor-independent
$\left( P_a^* -P_a \right)$
spin splitting $\Delta$ which violates heavy quark
spin flavor symmetry.
};
$\lambda_a$ describes wavefunction renormalization,
$$
Z_a = 1 + \lambda_a \left( M_K^2 / 16 \pi^2 f^2 \right)
\ln \left( M_K^2 / \mu^2 \right) ;
$$
$\beta_a$ parametrizes the nonanalytic $m_s^{3/2}$
contribution;
$\gamma_a$ describes the nonanalytic chiral logarithmic
corrections from the one-loop graphs shown in
\fig\figtwo{One-loop diagrams which contribute to
the heavy meson masses.  Solid dots denote the
axial pion-heavy meson vertex proportional to $g$
contained in $\CL_v^0$.
Diagram (a) is responsible
for the non-analytic $m_s^{3/2}$ mass contribution
and for wavefunction renormalization coefficients
$\lambda_a$.
Graphs (b) and (c) yield the coefficients $\gamma_a$
which occur in the $m_s^2 \ln m_s$ non-analytic
contributions to the heavy meson masses.
Diagrams (b) are proportional to $a$, and diagrams
(c) are proportional to $\sigma$.
Graph (d) yields the coefficients
$\gamma_a$ which parameterize the
$\Delta m_s \ln m_s$ contribution to the heavy
meson masses.  This non-analytic contribution
violates both $SU(3)_V$ flavor symmetry and heavy
quark spin flavor symmetry, and yields the leading
contribution to the mass difference
$\left\{ \left( P_s^* - P_s \right)
- \left( P^* - P \right) \right\}$.
};
$c_a$ are contact terms for the spin-independent
contributions to the masses arising from $\CL_v^{M M}$;
and $\delta c_a$ are contact terms for the hyperfine
splittings arising from $\CL_v^{M \s }$.  Only the
kaon mass $M_K$ appears in eqs.~\massave\
and~\massdiff\ since the pion mass vanishes for
$m_u = m_d = 0$ and the $\eta$ mass can be eliminated
using the Gell-Mann--Okubo formula
$M_\eta^2 = \frac 4 3 M_K^2$.
The scale $\mu$ is an arbitrary renormalization scale,
and is chosen to be $\mu \sim 1$~GeV.  The one-loop
expressions eqs.~\massave\ and~\massdiff\ including the
contact terms are independent of the scale $\mu$,
because the $\mu$ dependence of the chiral logarithms
is canceled by the $\mu$ dependence of the counterterms.
Changing the scale $\mu$ appearing in the chiral
logarithms redefines these local contact terms.
The chiral expansion
for the linear combination
$\frac 1 4 \left( P_a+ 3P^*_a \right)$ to one-loop order
including contact counterterms includes all
contributions to order $m_s^2$
since the tree-level mass splittings are linear in $m_s$.
In contrast, the one-loop calculation of the hyperfine
mass differences $\left( P^*_a- P_a \right)$ only
includes effects to
order $m_s$ since the
leading contribution $\Delta$ to these mass differences
is independent of $m_s$.  Calculation of the hyperfine
mass differences to order $m_s^2$ involves a two-loop
renormalization of the tree-level contribution $\Delta$,
a one-loop renormalization of the contributions
$\hyper \sigma m_s$ and $\hyper a m_s$, and contact
terms arising from the spin-dependent operators contained
in $\CL_v^{M M \s}$.

The coefficients appearing in eqs.~\massave\ and~\massdiff\
are calculated in terms of the parameters given in
Sect.~2.  The tree-level coefficients are
\eqn\tree{
\alpha_a = 2 \sigma m_s + 2  a m_s \delta_{3 a} .
}
The coefficients of the $m_s^{3/ 2}$ contribution\foot{The
$m_s^{3/2}$ contribution given here differs from
the correction
given in ref.~\goity\ by an overall factor of $\frac 4 3$.
The coefficients $\beta_a$ are simply related to
wavefunction renormalization coefficients $\lambda_a$ in
the same manner as for baryons \j \jmtwo\ since the spin
of the light degrees of freedom, which couples to pions,
 is $\half$ for both the
baryons and the heavy mesons.  In the case of the baryons
which contain no heavy quarks, the spin of the light degrees
of freedom is the total spin of the baryon. }
to the masses,
\eqn\loopb{\eqalign{
&\beta_{1,2} = g^2 \left( 1 + \frac 8 {3\sqrt 3} \frac 1 6
\right) \cr
&\beta_3 = g^2 \left( 2 + \frac 8 {3\sqrt 3} \frac 2 3
\right) ,\cr
}}
are obtained from the graph fig. 2(a).  This graph also
is responsible for the wavefunction renormalization
coefficients
\eqn\wf{\eqalign{
&\lambda_{1,2} = \frac 3 2 g^2 \left( \frac {11}9 \right)
\cr
&\lambda_3 = \frac 3 2 g^2 \left( \frac {26}9 \right) \ .
\cr
}}
The coefficient of the chiral logarithmic correction in
eq.~\massave\ is given by
\eqn\coefave{
\left( \gamma_a -\lambda_a \alpha_a \right)=
\cases{
-2  a m_s \left( \frac 1 2 - \frac 3 2 g^2 \right)
-2 \sigma m_s \left( \frac {26}9 \right)&$a=1,2$,\cr
\noalign{\medskip}
-2 a m_s \left( \frac {17} 9 + 3 g^2 \right)
-2 \sigma m_s \left( \frac {26}9 \right)&$a=3$,\cr}
}
while the counterterms obtained from $\CL_v^{M M}$
and $\CL_v^{M M \s}$ are
\eqn\ctave{
c_a = 4 d m_s^2 + \left( 4 c m_s^2 + 4  b m_s^2 \right)
\delta_{3 a} \ . }
The one-loop chiral logarithmic renormalization
of $\Delta$ is determined by
\eqn\coefdiff{
\left( \gamma_a -\lambda_a \Delta \right)=
- \Delta \left( \frac 2 3 \lambda_a \right)\ ,
}
whereas the counterterm\foot{Note that the $(a)$
superscript on $\hyper a$ is not an $SU(3)_V$ index!}
for this chiral logarithm is
\eqn\ctdiff{
\delta c_a= 2 \hyper\sigma  m_s + 2 \hyper a m_s \delta_{3 a}.
}

In the isospin limit, there are
four heavy meson masses $P$, $P^*$, $P_s$, and $P_s^*$. These
masses are determined by four linear combinations:  an $SU(3)_V$
symmetric, heavy quark spin-independent contribution
$\frac 1 4 \left( P + 3 P^* \right)$; an $SU(3)_V$-breaking,
spin-independent contribution
$\left\{ \frac 1 4 \left( P_s + 3 P_s^* \right)
-\frac 1 4 \left( P + 3 P^* \right) \right\}$;
an $SU(3)_V$ symmetric, heavy quark spin-dependent
contribution $\left( P^* - P \right)$; and an
$SU(3)_V$-violating, spin-dependent contribution
$\left\{ \left( P_s^* - P_s \right) -
\left( P^* - P \right) \right\}$.  These four linear
combinations are given by

\eqn\paramsa{\eqalign{
&\frac 1 4 \left( P + 3 P^* \right)  = m_H + 2  \sigma m_s
- g^2 \left( 1 + \frac 8 {3 \sqrt 3} \frac 1 6
\right) {{M_K^3} \over {16 \pi f^2}}
\cr
&\quad\qquad - 2  \sigma m_s \left(
2 \right)
{M_K^2 \over {16 \pi^2 f^2}} \ln \left( M_K^2/ \mu^2 \right)
\cr
&\quad\qquad - 2  \sigma m_s \left(
\frac23 \right)
{M_\eta^2 \over {16 \pi^2 f^2}} \ln \left( M_\eta^2/ \mu^2 \right)
\cr
&\quad\qquad-2 a m_s \left( \frac 1 2 - \frac 3 2 g^2 \right)
{M_K^2 \over {16 \pi^2 f^2}} \ln \left( M_K^2/ \mu^2
\right) + 4 d m_s^2 \cr
}}
\bigskip
\eqn\paramsb{\eqalign{
&\left\{ \frac 1 4 \left( P_s + 3 P_s^* \right)
-\frac 1 4 \left( P + 3 P^* \right) \right\} = 2  a m_s
- g^2 \left( 1 + \frac 8 {3 \sqrt 3} \frac 1 2
\right) {{M_K^3} \over {16 \pi f^2}}
\cr
&\quad\qquad- 2  a m_s \left( \frac12 + \frac 9 2 g^2 \right)
{M_K^2 \over {16 \pi^2 f^2}} \ln \left( M_K^2/ \mu^2
\right)
\cr
&\quad\qquad- 2  a m_s \left( \frac23 \right)
{M_\eta^2 \over {16 \pi^2 f^2}} \ln \left( M_\eta^2/ \mu^2
\right)
+4 c m_s^2+4 b m_s^2
\cr
}}
\bigskip
\eqn\paramsc{\eqalign{
&\left( P^* - P \right) = \Delta
-  g^2 \Delta
{M_K^2 \over {16 \pi^2 f^2}}
\ln \left( M_K^2/ \mu^2 \right)
\cr
&\quad- \frac16 g^2 \Delta
{M_\eta^2 \over {16 \pi^2 f^2}}
\ln \left( M_\eta^2/ \mu^2 \right)
+ 2 \hyper\sigma m_s
}}
\bigskip
\eqn\paramsd{\eqalign{
&\left\{ \left( P_s^* - P_s \right) -
\left( P^* - P \right) \right\}  =
-  g^2 \Delta
{M_K^2 \over {16 \pi^2 f^2}}
\ln \left( M_K^2/ \mu^2 \right)
\cr
&\quad- \frac12 g^2 \Delta
{M_\eta^2 \over {16 \pi^2 f^2}}
\ln \left( M_\eta^2/ \mu^2 \right)
+ 2 \hyper a m_s \ .
}}
In the above equations,  the chiral logarithms
$M^2_\eta\ln M_\eta^2/\mu^2$ have not been
rewritten in terms of $M_K$ in order to
define the
counterterms in a manner consistent
with the counterterms which appear in
equations derived for the
isospin splittings in the next section.

\newsec{Calculation of Heavy Meson Masses
Including $m_u$ and $m_d$}

In order to compute the isospin splittings amongst
the heavy mesons, the calculation of Sect.~3 must
be generalized to include nonvanishing $u$ and $d$
quark masses.  In addition
to isospin splittings due to the $m_d - m_u$ mass
difference,  there are isospin splittings due to
the difference of electromagnetic charges of the $d$
and $u$ quarks.
Electromagnetic
effects are expected to contribute significantly
to the isospin splittings of the heavy mesons,
and must be included.  In this section, the formul\ae\
of the previous section are  generalized to
include nonzero $u$ and $d$ quark masses, and electromagnetic
effects.

The appropriate generalizations of
eqs.~\massave\ and \massdiff\ to include nonzero
masses of the $u$ and $d$ quarks are given by
\eqn\nmassave{\eqalign{
\frac 1 4 \left( P_a + 3 P_a^* \right)&= m_H + \alpha_a
- \sum_{X=\pi,K,\eta}
\beta^{(X)}_a { M_X^3 \over {16 \pi f^2 }} \cr
&+ \sum_{X=\pi,K,\eta}
\left( \gamma_a^{(X)} - \lambda^{(X)}_a\, \alpha_a \right)
{M_X^2 \over {16 \pi^2 f^2}} \ln \left( M_X^2 / \mu^2
\right)
+ c_a \cr
}}
and
\eqn\nmassdiff{
\left( P^*_a - P_a \right)= \Delta
\quad + \sum_{X=\pi,K,\eta}
\left( \gamma_a^{(X)} - \lambda^{(X)}_a\, \Delta \right)
{M_X^2 \over {16 \pi^2 f^2}} \ln \left( M_X^2 / \mu^2
\right)
+ \delta c_a \ ,
}
where the sum over pseudo-Goldstone bosons $X$
now includes massive pions, and isospin splittings
in the pseudo-Goldstone boson masses are
retained.
The coefficients appearing in eqs.~\nmassave\
and~\nmassdiff\ are given in the appendix. Isospin splittings are very
small, as are mass effects proportional to the $u$ and $d$ quark masses.
We will therefore neglect terms which are of order $m_u(m_d-m_u)$, and
$m_d(m_d-m_u)$ in comparison to effects of order $m_s (m_d - m_u)$.
Finally, note that it is
important to include $\pi^0 - \eta$
mixing in the calculation of loop diagrams, since
the mixing angle is proportional to $(m_d - m_u)$.
Formul\ae\ relevant for the inclusion of this effect are given
in the appendix.

There are two isospin splittings,
a spin-independent splitting and a spin-dependent
splitting.  These splittings are determined from
eqs.~\nmassave\ and~\nmassdiff\ and from
electromagnetic splittings given in Sect.~2.
The spin-independent splitting is given by
\eqn\isorelnsa{\eqalign{
&\left\{ \frac 1 4 \left( P_d + 3 P_d^* \right)
- \frac 1 4 \left( P_u + 3 P_u^* \right) \right\} =
2  a (m_d - m_u)
- g^2 {{(M_{K^0}^3 - M_{K^-}^3) } \over {16 \pi f^2}}
\cr
&- g^2\sqrt{\frac43}\sin\phi{M^3_\eta \over {16 \pi f^2}}
- 2  a (m_d - m_u)
\left( \frac 1 2 + \frac 3 2 g^2 \right)
{M_K^2 \over {16 \pi^2 f^2}} \ln \left( M_K^2/ \mu^2
\right)\cr
&-2am_s\left(\frac12-\frac32 g^2\right)
\left[ {M_{K^0}^2 \over {16 \pi^2 f^2}}
\ln \left( M_{K^0}^2/ \mu^2 \right)
-{M_{K^-}^2 \over {16 \pi^2 f^2}}
\ln \left( M_{K^-}^2/ \mu^2 \right)\right]
\cr
&- 2 a (m_d - m_u)
\left( \frac 1 6 \right)
{M_\eta^2 \over {16 \pi^2 f^2}}
\ln \left( M_\eta^2/ \mu^2 \right)
+4 c (m_d - m_u) m_s
\cr
&-a_{em} \aem q_Q - \frac13 b_{em} \aem\cr
}}
and the hyperfine splitting is given by
\eqn\isorelnsb{\eqalign{
\left\{ \left( P_d^* - P_d \right) -
\left( P_u^* - P_u \right) \right\} &=
- g^2 \Delta
\left[ {M_{K^0}^2 \over {16 \pi^2 f^2}}
\ln \left( M_{K^0}^2/ \mu^2 \right)
-{M_{K^-}^2 \over {16 \pi^2 f^2}}
\ln \left( M_{K^-}^2/ \mu^2 \right)\right]
\cr
&- g^2 \Delta
\sqrt{\frac43}\sin\phi\left[ {M^2_\eta \over {16 \pi^2 f^2}}
\ln \left( M_{\eta}^2/ \mu^2 \right)
\right]
\cr
&+ 2 \hyper a (m_d - m_u)
-\hyperem a \aem q_Q -\frac13 \hyperem b \aem ,\cr
}}
where $\phi$ is the $\pi^0-\eta$ mixing angle
as defined in the appendix.
In eqs.~\isorelnsa\ and \isorelnsb, only terms of first order in
$(m_d-m_u)$ have been retained, and terms proportional to
$m_{u,d}(m_d - m_u)$ have been neglected compared to
terms proportional to $m_s(m_d - m_u)$. There is a consistency check
between the equations eq.~\paramsd\ and
eqs.~\isorelnsb. The $\mu$ dependence of the $\hyper a$ counterterm
is the same in both equations.

\newsec{Comparison with Experiment}

\nref\pdg{Particle Data Group, \physrev {D45}{1992} }
\nref\cleo{The Cleo Collaboration (D. Bortoletto {\it et al.})
Phys. Rev. Lett. 69 (1992) 2046 }
\nref\cusb{The CUSB-II Collaboration (J. Lee-Franzini
{\it et al.}), \prl {65}{1990}{2947} }
\nref\cleotwo{The Cleo II Collaboration (D.S. Akerib
{\it et al.}), \prl {67}{1991}{1692} }

The experimentally measured mass splittings
of mesons containing a single heavy quark are
\eqn\exptd{\eqalign{
&\left( D_s^+ - D^+ \right) =
99.5 \pm 0.6 {\rm\  MeV}\ \  \pdg
\cr
&\left( D^+ - D^0 \right) =
4.80 \pm 0.10 \pm 0.06 {\rm\  MeV}\ \ \cleo
\cr
&\left( D^{*+} - D^{*0} \right) =
3.32 \pm 0.08 \pm 0.05 {\rm\  MeV}\ \ \cleo
\cr
&\left( D^{*0} - D^{0} \right) =
142.12 \pm 0.05 \pm 0.05 {\rm\  MeV}\ \ \cleo
\cr
&\left( D^{*+} - D^{+} \right) =
140.64 \pm 0.08 \pm 0.06 {\rm\  MeV}\ \ \cleo
\cr
&\left( D_s^{*+} - D_s^{+} \right) =
141.5 \pm 1.9 {\rm\  MeV}\ \ \pdg
\cr
&\left\{ \left( D^{*0} - D^0 \right) -
\left( D^{*+} - D^+ \right) \right\} =
1.48 \pm 0.09 \pm 0.05 {\rm\  MeV}\ \ \cleo
\cr
}}
for the $D$ mesons, and
\eqn\exptb{\eqalign{
&\left( B_s - B \right) =
82.5 \pm 2.5 {\rm\  MeV}\ {\rm or}\
121 \pm 9 {\rm\  MeV}\ \ \cusb
\cr
&\left( B^0 - B^+ \right) =
0.01 \pm 0.08 {\rm\  MeV}\ \ \pdg
\cr
&\left( B^{*} - B \right) =
46.2 \pm 0.3 \pm 0.8 {\rm\  MeV}\ \ \cleotwo
\ {\rm or}\  45.4 \pm 1.0 {\rm\  MeV}
\ \ \cusb
\cr
&\left( B_s^{*} - B_s \right) =
47.0 \pm 2.6 {\rm\  MeV}\ \ \cusb
\cr
}}
for the $B$ mesons. These values can now be
compared with the theoretical formul\ae\
eqs.~\paramsa--\paramsd, \isorelnsa,
and~\isorelnsb.
In some instances, only qualitative statements
about the mass splittings can be made
because of unknown counterterm
parameters. We first discuss the spin-independent
splittings, and then the hyperfine splittings.
\bigskip
\leftline{\sl SPIN-INDEPENDENT SPLITTINGS}
\smallskip
We begin by comparing the experimental data
with the theoretical formula for the
$SU(3)_V$-violating mass splitting
$\left\{ \frac 1 4 \left( P_s + 3 P_s^* \right)
- \frac 1 4 \left( P_d + 3 P_d^* \right) \right\}$
given in eq.~\paramsb.  Only the pseudoscalar
splittings $(D_s^+ - D^+)$ and $(B_s - B)$ are
measured.  These pseudoscalar splittings are expected
to be equal to the
splitting $\left\{ \frac 1 4 \left( P_s + 3 P_s^* \right)
- \frac 1 4 \left( P_d + 3 P_d^* \right) \right\}$
to an accuracy of a few MeV, since $SU(3)_V$ violation in the
hyperfine splittings is only of this magnitude.
The theoretical formula given in eq.~\paramsb\
implies that the $s-d$ splitting has a contribution
which begins at zeroth order in the $1/ m_Q$ expansion,
${\it i.e.}$ the power series expansion in $1/ m_Q$
of the coefficient function $a$ begins at $\CO(1)$.
If $1/ m_Q$ effects are small relative to effects of
order unity, then the $s-d$ splitting will be
largely independent of the mass of the heavy quark.
Experimentally, the $(D_s^+ - D^+)$ and $(B_s - B)$
are comparable with each of magnitude about $100$~MeV.
The amount of disagreement of the experimental values
for the $D$ and $B$ systems is a measure of
$1/ m_Q$ contributions.  Experimentally, the $SU(3)_V$
splittings of the $D$ and $B$ mesons differ by roughly
$10$~MeV (or smaller), which is about a $10\%$ deviation.
The theoretical formula for the $SU(3)_V$ splitting
has a significant contribution from the $M_K^3$ term
which is independent of the splitting parameter $a$.
The $M_K^3$ term gives a negative contribution to the
splitting which can be as big as $-247$~MeV for $g^2
\ltap 0.5$.  The chiral logarithmic correction
acts in the opposite direction; for $\mu = 1$~GeV and
$g^2 \ltap 0.5$, the
parameter $2 a m_s$ gets a correction
$\ltap 0.9$ times its tree level value.  Thus, the
one-loop value of $2 a m_s$ can be significantly
greater than the value determined at tree-level of
approximately $100$~MeV. In addition, there are two
counterterm parameters $b$ and $c$ which cancel the $\mu$
dependence of the chiral logarithm.

The $s-d$ splitting can be compared with the $d-u$
isospin splitting eq.~\isorelnsa\ which depends on
the same parameter $a$. The central value of the
isospin splitting
$\left\{ \frac 1 4 \left( P_d + 3 P_d^* \right)
- \frac 1 4 \left( P_u + 3 P_u^* \right) \right\}$
is $3.69$~MeV for the $D$ mesons.
Only the pseudoscalar isospin splitting is measured for
the $B$ mesons; the experimental value is consistent with
zero and bounded to be less than a fraction of an
MeV.  Naively, one expects the isospin splitting
to be equal to the $s-d$ splitting times a suppression
factor of $(m_d - m_u)/ m_s \approx 1 /44$.
For a $s-d$ splitting of $100$~MeV, this scaling predicts
a $d-u$ splitting of $2.3$~MeV, which is about the
right magnitude.
Because the isospin splitting is so small,  heavy quark
flavor violating and electromagnetic
effects which are irrelevant for the
$s-d$ splitting can be significant for the isospin
splitting. The isospin splittings of
the $D$ and $B$ mesons are expected to differ by
approximately $2-3$~MeV due to electromagnetic effects,
eq.~\lqbigQ. In addition, the two counterterms $b$ and $c$
relevant
for the $s-d$ and isospin splittings have different
scaling dependence on the light quark masses.  Thus,
a more accurate prediction of the isospin splitting from
the $s-d$ splitting can not be
expected.
Evaluation of the $M^3$ contribution
contributing to the isospin splitting in eq.~\isorelnsa\
leads to a negative contribution which can be as
large as $5.6$~MeV for $g^2 \ltap 0.5$.  The chiral
logarithmic corrections to
$2 a (m_d - m_u)$ are about 0.3 times the tree level value.
All these terms are comparable in size to the measured isospin
splitting.
In summary, the $s-d$ and $d-u$ spin independent splittings are
consistent with experiment, but definite predictions cannot
be made because of the unknown counterterms $b$ and $c$.
\bigskip
\leftline{\sl HYPERFINE SPLITTINGS}
\smallskip
The $SU(3)_V$ invariant hyperfine splittings of the
$D$ and $B$ meson are effects which start at order
$1/ m_Q$ in the heavy quark mass expansion.
The hyperfine splitting for
the $D$ mesons is measured to be about $140$~MeV, whereas
the hyperfine splitting is about $46$~MeV for the
$B$ mesons.  The ratio of these two splittings is
consistent with what one would expect from naive
$1/ m_Q$ scaling.  The theoretical formul\ae\ for
the hyperfine splitting $P^* - P$ in eq.~\paramsd\
predicts a chiral logarithmic correction of
the tree-level parameter $\Delta$
of $\ltap 0.15 \Delta$
for $g^2 \ltap 0.5$.

The theoretical formul\ae\
eqs.~\paramsa--\paramsd,\isorelnsa\ and~\isorelnsb\
for $SU(3)_V$-breaking hyperfine splittings
depend on only two
parameters, $\Delta$ and a single counterterm
proportional to $\hyper a$.
The leading contribution
to the $SU(3)_V$ violation in the hyperfine splittings
is a chiral logarithmic correction of $\Delta$.
For the $s-d$ hyperfine splitting, this correction
is $\ltap 0.2 \Delta\approx 28$~MeV for $g^2 \ltap 0.5$.
The total $s-d$ hyperfine splitting is of order 1~MeV, so
the counterterm $2\hyper a m_s$ must be chosen to be $-27$~MeV
for a nearly complete cancellation of the chiral logarithm.
For the isospin violating $d-u$ hyperfine splitting, the chiral
logarithmic correction is
$\ltap 2 \times 10^{-3}\Delta \approx0.28$~MeV, and
accounts for only a fraction of the hyperfine splitting.
Thus most of the isospin hyperfine splitting comes from
the counterterm and electromagnetic corrections.
The $\hyper a$ counterterm contribution to the isospin hyperfine
splitting is $(m_d-m_u)/m_s$ times $-27$~MeV, or $-0.6$~MeV.
Together, the chiral logarithm and the counterterm
account for only approximately $-0.3$~MeV of the
isospin hyperfine splitting.
The measured value of the isospin
hyperfine splitting
$\left\{ \left( \left( D^{*+} - D^+ \right)
-D^{*0} - D^0 \right) \right\}=-1.48\pm0.09\pm0.05$
therefore determines the electromagnetic contribution to the isospin
hyperfine splitting to be about $-1.2$~MeV, which is of the
order of magnitude expected for electromagnetic hyperfine
splittings. The isospin hyperfine splitting for the
$B$ mesons has not been measured. The $1/m_Q$ scaling
of $\Delta$ and $\hyper a$ implies
that the contribution to the $B$ meson hyperfine splittings
from the $d - u$ mass difference is
of order $-0.3 (m_D/m_B)=-0.1$~MeV. The electromagnetic
contribution, however, does
not scale like $m_D/m_B$ because of flavor symmetry violation in
$\hyperem a$ from the heavy quark charge.

\newsec{Conclusions}

The experimental data for $D$ and $B$ meson $SU(3)_V$
and hyperfine splittings is consistent with the
theoretical calculation to one-loop order in chiral
perturbation theory.  The $1/ m_Q$ corrections
to the leading predictions for $SU(3)_V$ splittings
are found to be a small correction experimentally.
Hyperfine
splittings, whose leading dependence on $1 / m_Q$
begins at linear order in the $1 / m_Q$ expansion,
are found to satisfy naive $1/ m_Q$ scaling
experimentally.  Thus, leading order scaling in the
the heavy quark mass as implied by the heavy quark
mass expansion is evident in the experimental data.
Naive $1/ m_Q$ scaling implies that
\eqn\massrelns{
\frac 1 4 \left\{ \left( D_s + 3 D_s^* \right)
-\left( D + 3 D^* \right) \right\}
=\frac 1 4 \left\{ \left( B_s + 3 B_s^* \right)
-\left( B + 3 B^* \right) \right\}
}
up to corrections of order $1/ m_Q$ and that
\eqn\massrelnstwo{
{ {\left( B^* - B \right) }
\over {\left( D^* - D \right) }}
= { {\left\{ \left( B_s^* - B_s \right) -
\left( B^* - B \right) \right\} } \over
{\left\{ \left( D_s^* - D_s \right) -
\left( D^* - D \right) \right\} }}.
}
The two ratios above equal $m_c / m_b$ to leading
order in $m_Q$.  Generalizations of eqs.~\massrelns\
and~\massrelnstwo\ to the isospin splittings are
violated by electromagnetic effects which are
significant.  Finally,
there are a large number of counterterms in the heavy
meson lagrangian at one loop, so very few
definite predictions which are independent of the values
of the counterterms can be obtained.
The experimental data can be used
to determine the one-loop counterterms once the value of
$g^2$ is measured.

The most important conclusion of this work, however,
is that the $s-d$ and isospin splittings of $D$ and
$B$ mesons are consistent with the theory.  No breakdown
of chiral perturbation theory or the heavy quark mass
expansion is in evidence from the theoretical analysis
of the $D$ and $B$ meson masses.

\vfill\break\eject

\centerline{\bf Acknowledgements}

I would like to thank M.~Luke, A.V.~Manohar, A.~Pich,
J.L.~Rosner, M.J.~Savage and M.B.~Wise for discussions.
I would like to thank M.B.~Wise for informing me that he and
L.~Randall are also studying the non-analytic corrections to
the heavy meson masses.

\appendix{A}{Coefficients for Chiral Expansion in
Light Quark Masses}
\noindent

\eqn\nmassavetwo{\eqalign{
\frac 1 4 \left( P_a + 3 P_a^* \right)&= m_H + \alpha_a
- \sum_{X=\pi,K,\eta}
\beta^{(X)}_a { M_X^3 \over {16 \pi f^2 }} \cr
&+ \sum_{X=\pi,K,\eta}
\left( \gamma_a^{(X)} - \lambda^{(X)}_a\, \alpha_a \right)
{M_X^2 \over {16 \pi^2 f^2}} \ln \left( M_X^2 / \mu^2
\right)
+ c_a \cr
}}

\noindent
Tree-level coefficients:
\eqn\ntree{
\alpha_a = 2  \sigma (m_u +m_d +m_s)
+ 2  a (m_u \delta_{1 a}
+ m_d \delta_{2 a} + m_s \delta_{3 a})
}

\noindent
$m_q^{3 / 2}$ coefficients (including isospin violation for $a=1,2$):
\eqn\nloopb{
\eqalign{
&\beta^{(\pi^0)}_{1} = g^2\left[\frac1{\sqrt2}c_\phi+
\frac1{\sqrt6}s_\phi\right]^2\cr
&\beta^{(\pi^-)}_{1} = g^2 \cr
&\beta^{(K^-)}_1 = g^2\cr
&\beta^{(\eta)}_{1} = g^2\left[\frac1{\sqrt6}c_\phi-\frac1{\sqrt2}
s_\phi\right]^2 \cr
}
\qquad\eqalign{
&\beta^{(\pi^0)}_{2} = g^2\left[-\frac1{\sqrt2}c_\phi+
\frac1{\sqrt6}s_\phi\right]^2 \cr
&\beta^{(\pi^+)}_{2} = g^2 \cr
&\beta^{(K^0)}_2 = g^2 \cr
&\beta^{(\eta)}_{2} = g^2\left[\frac1{\sqrt6}c_\phi+\frac1{\sqrt2}
s_\phi\right]^2 \cr
}
\qquad\eqalign{
&\beta^{(\pi)}_3 = 0 \cr
&\beta^{(K)}_3 = 2 g^2 \cr
&\beta^{(\eta)}_3 = \frac 2 3 g^2 \cr
}}
Here $s_\phi$ and $c_\phi$ are the sine and cosine of the $\pi^0-\eta$
mixing angle,
where the $\pi^0$ and $\eta$ fields in eq.~\pion\ are related to the
mass eigenstate fields $\pi^0_{\rm phys}$ and $\eta_{\rm phys}$ by
$$\eqalign{
\pi^0_{\rm phys} &= \cos\phi\ \pi^0 +\sin\phi\ \eta,\cr
\eta_{\rm phys} &=  -\sin\phi\ \pi^0 + \cos\phi\ \eta.\cr
}$$
The $\pi^0-\eta$ mixing angle is determined to first order in isospin
breaking to be
$$
\sin\phi\approx\phi= {\sqrt3\left(m_d-m_u\right)\over 4 m_s}.
$$

\noindent
Wavefunction renormalization coefficients:
\eqn\nwf{
\lambda^{(X)}_a = \frac 3 2 \beta^{(X)}_a
}

\noindent
One-loop coefficients:
\eqn\ncoefpi{\eqalign{
\left( \gamma_a^{(\pi)} -\lambda_a^{(\pi)} \alpha_a \right)&=
-  a \left[ (2 m_u + m_d) \delta_{1 a}
+(m_u + 2 m_d) \delta_{2 a} \right] \cr
&+ 3 g^2  a
\left[ (- m_u + m_d) \delta_{1 a}
+ (m_u -  m_d ) \delta_{2 a} \right]
\cr
&- 3 \sigma (m_u + m_d)
\cr
}}

\eqn\ncoefkp{\eqalign{
\left( \gamma_a^{(K^+)} -\lambda_a^{(K^+)} \alpha_a \right)&=
-  a
\left[ (m_u + m_s) \delta_{1 a}
+(m_u + m_s) \delta_{3 a} \right] \cr
&+ 3 g^2  a
\left[ (m_s - m_u) \delta_{1 a}
+(m_u -m_s) \delta_{3 a} \right] \cr
&-2  \sigma (m_u + m_s)
\cr
}}
\eqn\ncoefkz{\eqalign{
\left( \gamma_a^{(K^0)} -\lambda_a^{(K^0)} \alpha_a \right)&=
-  a
\left[ (m_d + m_s) \delta_{2 a}
+( m_d +  m_s) \delta_{3 a} \right] \cr
&+ 3 g^2  a
\left[ (m_s - m_d) \delta_{2 a}
+(m_d - m_s) \delta_{3 a} \right] \cr
&-2  \sigma ( m_d +  m_s)
\cr
}}

\eqn\ncoefeta{\eqalign{
\left( \gamma_a^{(\eta)} -\lambda_a^{(\eta)} \alpha_a \right)&=
-  a
\left[ \frac 1 3 m_u \delta_{1 a} + \frac 1 3 m_d \delta_{2 a}
+\frac 4 3  m_s \delta_{3 a} \right]
\cr
&-\frac 1 3  \sigma (m_u + m_d + 4 m_s)
\cr
}}

\noindent
Counterterm coefficients:
\eqn\nctave{\eqalign{
&c_a = 4 d (m_u^2 + m_d^2 + m_s^2)
+ 4 c (m_u + m_d + m_s)
(m_u \delta_{1 a} + m_d \delta_{2 a}
+ m_s \delta_{3 a}) \cr
&\qquad + 4  b (m_u^2 \delta_{1 a} + m_d^2 \delta_{2 a}
+ m_s^2 \delta_{3 a})
\cr
}}

\eqn\nmassdifftwo{
\left( P^*_a - P_a \right)= \Delta
\quad + \sum_{X=\pi,K,\eta}
\left( \gamma_a^{(X)} - \lambda^{(X)}_a\, \Delta \right)
{M_X^2 \over {16 \pi^2 f^2}} \ln \left( M_X^2 / \mu^2
\right)
+ \delta c_a
}

\noindent
One-loop coefficients:
\eqn\ncoefdiff{\eqalign{
&\left( \gamma_a^{(X)} -\lambda_a^{(X)} \Delta \right)=
-\Delta \left( \frac 2 3 \lambda_a^{(X)} \right)
=-\Delta\beta_a^{(X)} \cr
}}
using eq.~\nwf.

\noindent
Counterterm coefficients:
\eqn\nctdiff{
\delta c_a = 2 \hyper\sigma (m_u + m_d + m_s)
+2 \hyper a (m_u \delta_{1 a}
+ m_d \delta_{2 a} + m_s \delta_{3 a})
}

\listrefs
\listfigs
\vfill
\eject
\null\vskip 2.in
\insertfig{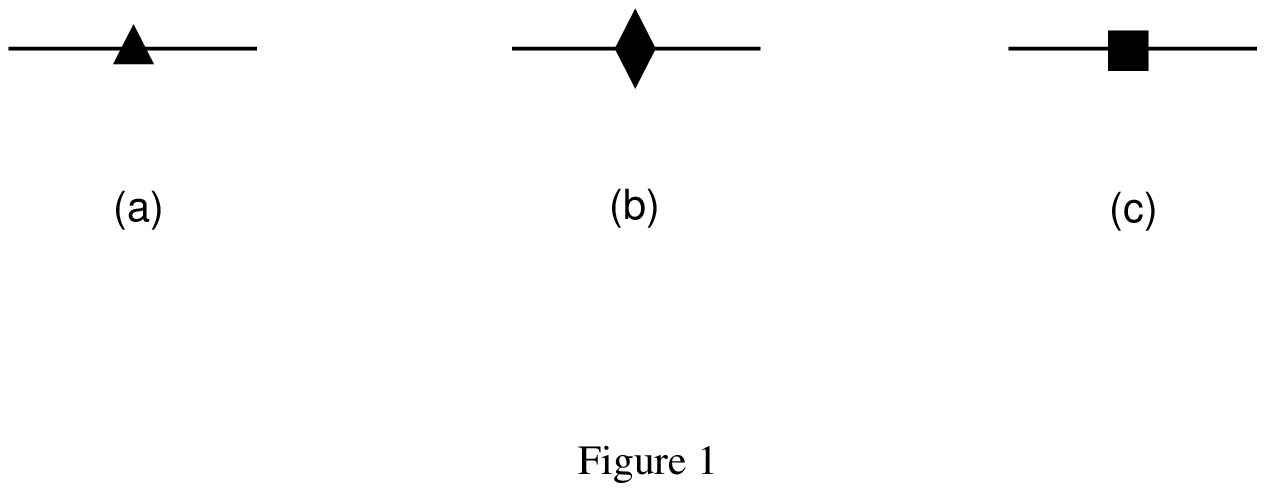}
\vfill
\eject
\null\vskip 1.in
\insertfig{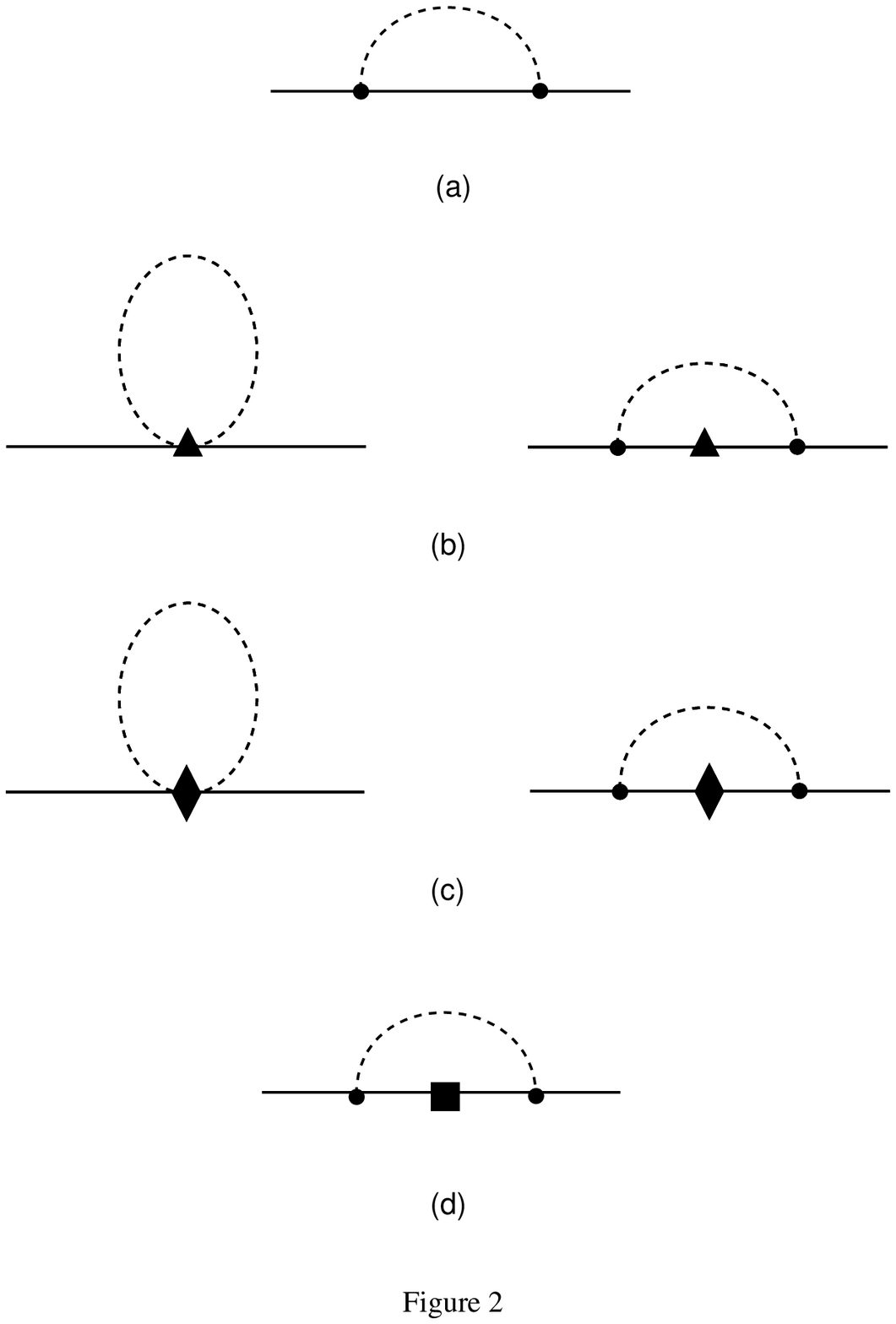}
\bye